\documentclass[a4paper, aps, prd, 11pt, showpacs, superscriptaddress, floatfix, nofootinbib, reprint]{revtex4-1}

\usepackage{geometry}
\usepackage[activate={true,nocompatibility},tracking=true,kerning=true,factor=1000,stretch=10,shrink=10]{microtype}
\usepackage[usenames]{color}
\usepackage{graphicx}
\usepackage{bm, amssymb, amsmath, amsfonts}
\usepackage{units,verbatim,subfigure}
\usepackage{multirow}
\usepackage{hyperref}
\hypersetup{
	colorlinks=true, 	% false: boxed links; true: colored links
	linkcolor=blue,		% color of internal links
    citecolor=cyan,		% color of links to bibliography
}

\def \be{\begin{equation}}
\def \ee{\end{equation}}
\def \bea{\begin{eqnarray}}
\def \eea{\end{eqnarray}}
\def \ba{\begin{eqnarray}}
\def \ea{\end{eqnarray}}

\def \Msun {M_\odot}

\def \Msun{M_{\odot}}

\newcommand{\RNs}{\mathbb{R}^{N_s}}
\newcommand{\Rl}{\mathbb{R}^{\ell}}
\DeclareMathOperator{\rank}{rank}
\newcommand{\iu}{{i\mkern1mu}}
\newcommand*{\rom}[1]{\expandafter\@slowromancap\romannumeral #1@}
\newcommand{\RNum}[1]{\uppercase\expandafter{\romannumeral #1\relax}}

\newcommand{\IUCAA}{Inter-University Centre for Astronomy and
  Astrophysics, Post Bag 4, Ganeshkhind, Pune 411 007, India}

\newcommand{\WSU}{Department of Physics \& Astronomy, Washington State University, 1245 Webster, Pullman, WA 99164-2814, U.S.A.}

\newcommand{\IITGN}{Indian Institute of Technology Gandhinagar, Gujarat 382355, India}

\newcommand{\IISERPUNE}{Indian Institute of Science Education and Research, Homi Bhabha Road, Pune 411008, India}

\newcommand{\IITK}{Department of Physics, Indian Institute of Technology, Kanpur 208016, India}

\newcommand{\IBM}{IBM Research, Bangalore 560045, India}

\begin{document}

%\linenumbers

\title{Random projections in gravitational wave searches of compact binaries 
%\\ \vspace{1em} {\color{blue}\textsc{In no particular order}}
}

\author{Sumeet Kulkarni}
%\email{sumeet.kulkarni@students.iiserpune.ac.in}
\affiliation{\IISERPUNE}

\author{Khun Sang Phukon}
%\email{khunsang@iitk.ac.in}
\affiliation{\IITK}

\author{Amit Reza}
%\email{amit.reza@iitgn.ac.in}
\affiliation{\IITGN}

\author{Sukanta Bose}
\affiliation{\IUCAA}
\affiliation{\WSU}

\author{Anirban Dasgupta}
\affiliation{\IITGN}

\author{Dilip Krishnaswamy}
\affiliation{\IBM}

\author{Anand S. Sengupta}
\affiliation{\IITGN}

%\date{\today}
%\pacs{04.30.-w, 07.05.Kf,95.75.Pq, 95.75.Wx}

\begin{abstract}
Random projection (RP) is a powerful dimension reduction technique widely used in analysis of high dimensional data. We demonstrate how this technique can be used to improve the computational efficiency of gravitational wave searches from compact binaries of neutron stars or black holes. Improvements in low-frequency response and bandwidth due to detector hardware upgrades pose a data analysis challenge in the advanced LIGO era as they result in increased redundancy in template databases and longer templates due to higher number of signal cycles in band. The RP-based methods presented here address both these issues within the same broad framework. We first use RP for an efficient, singular value decomposition inspired template matrix factorization and develop a geometric intuition for why this approach works. We then use RP to calculate approximate time-domain match correlations in a lower dimensional vector space. For searches over parameters corresponding to non-spinning binaries with a neutron star and a black hole, a combination of the two methods can reduce the total on-line computational cost by an order of magnitude over a nominal baseline. This can, in turn, help free-up computational resources needed to go beyond current spin-aligned searches to more complex ones involving generically spinning waveforms.

\end{abstract}

%\preprint{{\color{red}[LIGO-P1700443]}}

\maketitle
%--- Introduction
%\noindent{\textit{\textbf{Introduction.}}} 
\section{Introduction}

The direct detections of gravitational waves (GWs) from the mergers of black holes and neutron stars~\cite{Abbott:2016GW150914,Abbott:2016GW151226, TheLIGOScientific:2016pea,Abbott:20170104,Abbott:2017GW170814,gw170817} by Advanced LIGO (aLIGO)~\cite{advLigo2015} and Advanced Virgo (AdV)~\cite{advVirgo2015} detectors in the first and second observing runs (O1 and O2, respectively) have launched the era of GW astronomy~\cite{Monitor:2017mdv,GBM:2017lvd}. 
In the coming years, the global network of ground-based detectors,
comprising aLIGO, AdV, KAGRA~\cite{Akutsu:2017thy} and LIGO-India~\cite{Unnikrishnan:2013qwa} will 
%join the aLIGO and AdV detectors %to create a global network of second-generation GW observatories to 
not only increase the detection rate and 
%but also improve the localization of GW sources - thereby,
facilitate the search for their possible electromagnetic counterparts~\cite{Rana:2016crg,Srivastava:2016fyr,Ghosh:2013yda}
but also produce an unprecedentedly large amount of data, which can pose an interesting computational challenge for GW data analysis.

At present, 
%three different approaches: adiabatic, non-adiabatic and phenomenological, are employed to 
theoretically modeled
%% the expected GW signal from the inspiral, merger and ringdown phases of the evolution of compact binary systems. These 
compact binary coalescence (CBC) waveforms are used as templates to {\em matched-filter}~\cite{Helstrom:1994} the detector data in these searches~\cite{Owen:1995tm,Sathyaprakash:1991mt}.  
%The matched-filtering operation is implemented by cross-correlating the detector data with these templates. The template waveforms depend on a set of extrinsic parameters (e.g. the time $t_0$ and phase $\phi_0$ at  arrival or coalescence of the signal in band), and intrinsic parameters (e.g., component masses and their spins). The matched-filter output $\rho$, is maximized over these parameters to get the optimal detection statistic and is often referred to as the signal-to-noise ratio (SNR). While $\rho$ can be algebraically maximized over extrinsic parameters, 
%%a brute force approach is employed for intrinsic parameters, 
%it is evaluated over a large bank of templates constructed for a grid of values of the intrinsic parameters spanning astrophysical relevant ranges.
%A template bank is at first constructed by placing a suitable grid spanning astrophysical relevant range of the intrinsic parameters. Thereafter, $\rho$ is evaluated at each point in this grid which makes it a computationally expensive business. 
A brute force computation of this cross-correlation with a suitable grid of templates spanning astrophysical ranges of search parameters can be expensive (but see~\cite{Prix:2007ks,Harry:2009ea,Roy:2017qgg}).
% CHECK: Several different grid placement algorithms have been developed  to minimize the required number of grid points~\cite{Prix:2007ks,Harry:2009ea,Roy:2017qgg}. 
%These include lattice based methods~\cite{Prix:2007ks}, stochastic placement strategies~\cite{Harry:2009ea} and a hybrid combination of the two approaches~\cite{Roy:2017qgg}. 
As these detectors are paced through planned upgrades, one expects better sensitivity at low frequencies and an increase in the detector bandwidth. 
%The seismic cutoff frequency is expected to decrease from $30$ to $10$ Hz. 
The combined effects of these changes will not only increase the volume of the search parameter space but also result in denser template banks, thereby increasing their redundancy.
%The improvement in low frequency response of the detector will also lead to 
More cycles of the signal will fall in band and increasing their duration. These highlight the need for designing efficient and scalable methods for matched-filtering-based templated CBC searches~\cite{Cokelaer:2007kx,Abbott:2007ai,Harry:2013tca,Babak:2008rb,Manca:2009xw,Privitera:2013xza}.

In a seminal work, Cannon {\em et al.}~\cite{Cannon:singular,Cannon:2011xk,Cannon:2011vi} showed how singular value decomposition (SVD) can mitigate the redundancies in CBC template banks by effectively reducing the number of filters or templates, owing to their strong correlation for similar parameter values, with negligible effect on search performance. We show, however, that the  computational cost of SVD factorization does not scale favorably with increase in bank size. Further, it may not be possible to factorize very large banks {\it in toto} as it requires prohibitively large random access memory.

Random Projections (RP), conceived by the pioneering work of Johnson and Lindenstrauss~\cite{johnson1984extensions}, is a computationally efficient technique for dimension reduction and finds applications in many areas of data science~\cite{bingham2001random}. In this {\textit{Letter}}, we apply this technique to address two key challenges in future CBC searches: handling redundancies in large template databases; and efficiently correlating noisy data against long templates. 

The {\em primary impact} of this work is multi-fold: (1)~Efficient template matrix factorization can be used to address the {\em redundancy problem}. This is similar in spirit to the SVD factorization that is at the heart of the ``GstLAL"-based inspiral pipeline~\cite{Messick:2016aqy,Cannon:2011vi,Privitera:2013xza}, but our RP method scales well for very large number of templates embedded in high-dimensional Euclidean space. 
%In many scenarios 
Such factorizations can be done off-line, in advance of a CBC search. Nonetheless there can be situations when the factors need to be updated on-line, e.g., owing to the non-stationarity of data. Our adaptations will benefit both scenarios. (2) We show the explicit connection between the new factorization scheme and the extant SVD method. This bridges the two approaches and makes it readily usable. (3) The computational challenges arising from correlating noisy data against long templates (also known as the {\em curse of dimensionality}) is addressed by casting the match calculation in a lower-dimensional space. 
%For certain types of template banks (e.g., for CBCs with precessing spins), the template matrix may be less amenable to a SVD-like factorization. There the total computational cost can be significantly reduced by using the RP-based correlation alone. 
(4) Finally, we show that RP-based template matrix factorization and match computation in reduced dimension can be combined effectively for efficient CBC searches.

Currently the GstLAL-based inspiral pipeline utilizes time-slicing of templates to improve computational efficiency, and also involves spin-aligned templates. Since it is for the first time that the RP is being introduced in GW searches, our primary objective here is to elucidate how its core ideas can help them. This is why we demonstrate application of RP in the simple case of a single slice of data and non-spinning inspiral templates. This simplification notwithstanding, the RP-based methods introduced here can be readily applied to time-sliced data and spin-aligned templates. (A detailed study of that application and the computational advantage so gained will be presented in a future work.)

%-- State of the art of CBC searches
%\noindent{\textit{\textbf{Compact binary searches.}}} 
\section{Compact binary searches}

Consider a CBC search involving a bank of $N_T$ templates over a given parameter space. Following the convention in Ref.~\cite{Cannon:singular}, let $\bf H$ denote the $2N_T \times N_s$ {\textit{template matrix}} with $2N_T$ rows of real-valued unit-norm whitened %templates, (Note Canon et al. make a distinction between filters and templates. Each template has two filters.
filters, each sampled over $N_s$ time-points. The template matrix may be viewed as $2 N_T$ row-vectors embedded in $N_s$-dimensional Euclidean space $\mathbb{R}^{N_s}$.
The complex matched-filter output of the $\alpha^{\rm th}$ template at a specific point in time, against the whitened data $\vec{S}$ is the 
%vector 
inner product: 
\be
\label{snr_vec}
\rho_\alpha  =  \left ( H_{(2 \alpha -1)}  - \iu H_{(2\alpha)} \right ) %\cdot 
\vec{S}^T \,,
\ee
where $H_\alpha$ denotes the $\alpha^{\rm th}$ row of ${\bf H}$ and $\vec{S}^T$ is the transpose of $\vec{S}$. The signal-to-noise ratio maximized over the initial phase $\phi_0$, is given by $| \rho_\alpha | $.
%and $\mathbf{A}_\alpha$ is a matrix that, in general, can have complex entries
% This is wrong! \langle \mathbf{H}_\alpha, \vec{S} \rangle =  \mathbf{H}^\ast_\alpha \cdot \vec{S}^T 
% into this later: where, $\ast$ denotes complex conjugation and $\Delta t$ is the circular time shift of the template with respect to $\vec S$. 
In our notation, the ${H}_\alpha$'s and signal $\vec S$ are assumed to be row vectors. The overlap between two templates,
%${H}_\alpha$ and ${H}_\beta$ is 
%${H}_\alpha \cdot {H}^T_\beta$, 
when maximized over \textit{extrinsic} parameters
(e.g., the time $t_0$ and phase $\phi_0$ at  arrival or coalescence of the signal in band),
%DEFINED before, namely, $t_0$ and $\phi_0$, 
produces the \textit{match}. The match between templates with similar intrinsic parameters (such as the compact object masses and spins), 
can be very high - signifying the rank deficiency of the template matrix.
A typical {\textit{off-line}} 
CBC search involves calculating the cross-correlation between $\vec S$ and every row of $\mathbf{H}$ for a series of relative time-shifts, or values of $t_0$, thereby generating a time-series of $\rho_\alpha$ values, for every $\alpha$.
The use of a large number of templates ($N_T$), each sampled over a large number of points ($N_s$) amplifies the search's computational cost.

\begin{figure}[!htb]
\centering{
\includegraphics[width=.5\textwidth]{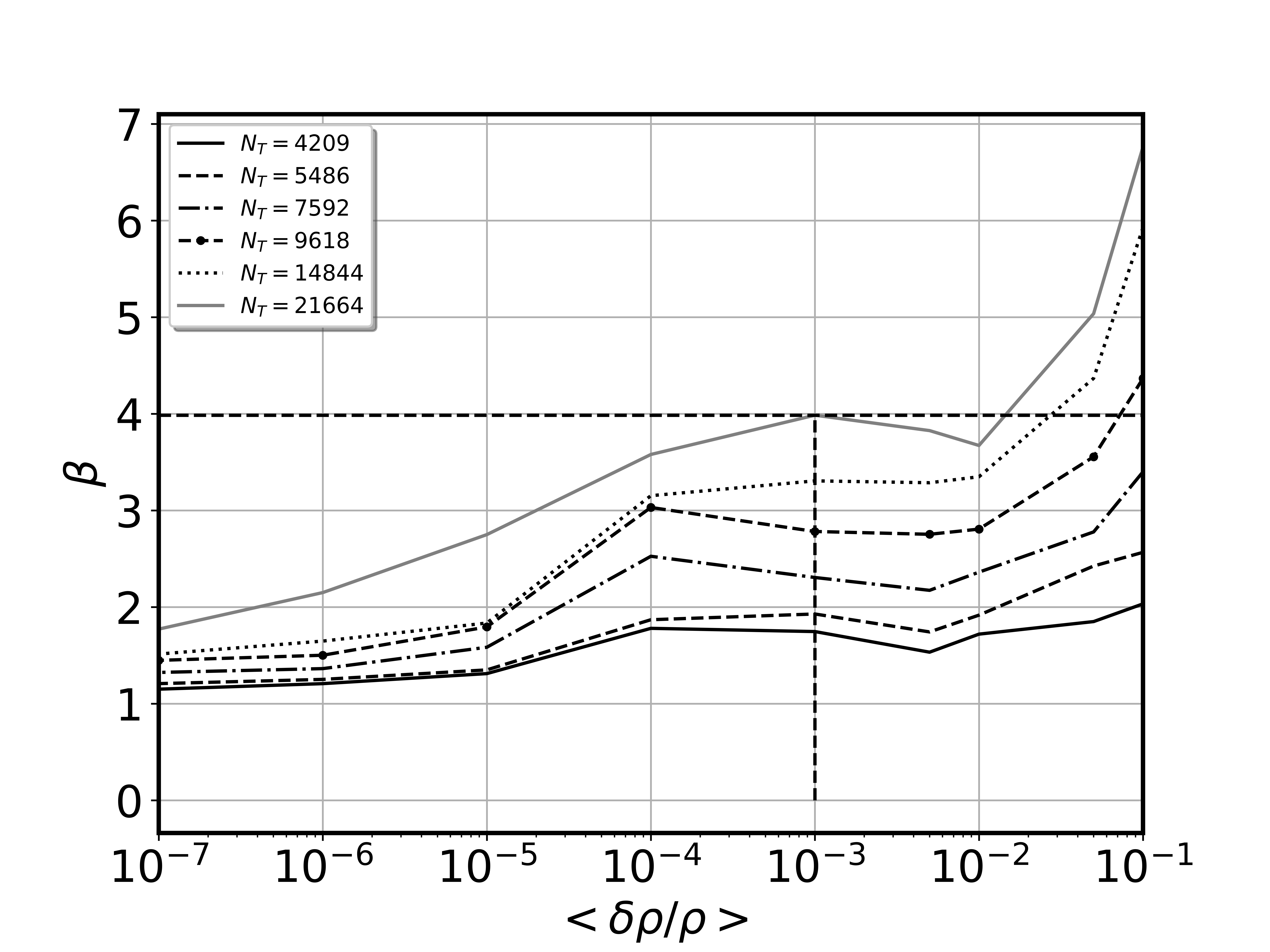}}
\caption{\label{Split_Comp} The factor by which the number of SVD basis vectors increases due to partitioning of template bank of size $N_T$ into sub-banks of $500$ templates each is shown as $\beta$ on the vertical axis.
%-- \beta is now defined in the text. No need to define it here again!
%$\beta$ is the ratio of total number of basis vectors obtained from SVD of sub-banks to the number of basis vectors from SVD of the full banks, at a fixed average loss in SNR reconstruction $<\delta\rho/\rho>$. 
Results from six different template-bank sizes are shown. For example,
the bank with 4209 templates is divided into eight sub-banks of 500
templates each and a ninth one of 209 templates. There $\beta$ reaches
a high of $\sim 2$ when one tolerates an average fractional loss in
SNR of $\langle\delta\rho/\rho\rangle \sim 0.1$. On the other hand,
for any of these template banks, as one approaches machine-precision accuracy in SNR reconstruction, $\beta \rightarrow 1$ as expected. A practical operating point would be $\langle\delta\rho/\rho\rangle \sim 10^{-3}$. The trend from the six examples shown here indicates that $\beta$ can be quite large for searches in aLIGO data where $N_T \sim 10^5$.}
\end{figure}

The rank deficiency of $\mathbf{H}$ is exploited in the truncated SVD approach, where every row is approximated as a linear combination of only  
% "top" may not be clear to some readers: the top-$\ell$ 
$\ell$ of the $2N_T$ right singular vectors with the most dominant
singular values. Further, these ``basis" vectors are used as
eigen-templates against which the data are cross-correlated. The left
singular vectors of $\mathbf{H}$ and the singular values are combined
into a coefficient matrix that is used to reconstruct the approximate
signal-to-noise ratio (SNR). The truncation of the basis leads to errors in the approximation of the template waveforms, which further translates to imperfect reconstructions of the SNR. The fractional SNR loss can be measured as a function of the discarded $(2N_T- \ell)$ singular values.  

The SVD factorization of the template matrix $\mathbf H $ has a time-complexity proportional to $\mathcal{O} (N_T^2 N_s)$ assuming $N_T \leq N_s$. Thus, such factorizations fast become computationally unviable with increasing 
%intractable for larger 
size of a template bank.
%due to the steep increase in computational cost. Moreover, 
Since the entire template matrix can become too large to be saved in single machine memory, a suitable parallel scheme is required to apply SVD to larger banks. SVD-based on-line CBC searches~\cite{Cannon:2011vi,Messick:2016aqy} work around this problem by splitting the bank into smaller sub-banks that are more amenable to such factorization separately.
%independent of one another. 
While the optimal way of partitioning the bank is an open problem, the act of splitting the bank prevents exploitation of the {\em linear dependency} of templates across the sub-banks.
% for computational savings.
%this leads to a loss of {\em linear dependency} of the templates across the sub-banks  which has important consequences in the computational cost of such searches. 
This is seen in Fig.~\ref{Split_Comp}, where we plot $\beta$, which is defined as the ratio of the number of basis vectors summed across all the sub-banks to the number of basis vectors from the SVD factorization of the full bank, at a given average fractional loss in accuracy of the reconstructed SNR. 
%It is clear that 
By splitting the bank, one effectively ends up requiring many more eigen-templates against which the data are filtered. 
%One can estimate the value of $\beta$ for a large template bank. 
When extrapolated to realistic template bank sizes of $N_T  \approx 10^6$, $\beta$ can be as large as 
%$20$ to 
$\sim 10^2$ at $\langle \frac{\delta \rho}{\rho} \rangle = 10^{-3}$.

%-- Already mentioned in the figure caption. No need to repeat!
%From the trends observed, it can be concluded that this factor can be quite large for template banks for detectors like Advanced LIGO (aLIGO) and Virgo, where $N_T \sim 10^5$.
%

The SVD-inspired RP-based 
%template matrix 
factorization presented below addresses this issue and is scalable for large template banks. 
We also apply RP to calculate the match correlations in a lower-dimensional space $\mathbb{R}^k$ where $k \leq N_s$. These correlations could be either between templates %and the data as shown in Eq.~(\ref{snr_vec}) 
or between basis vectors within the SVD paradigm. The full potential of the RP-based methods introduced here can be realized by combining them together. We demonstrate its feasibility with an example. 

%--- Introduce RP 
%\noindent{\textit{\textbf{Random projection.}}}
\section{Random projection}
\label{projection_random}

The core theoretical idea behind the RP technique is the  Johnson-Lindenstrauss (JL) lemma~\cite{johnson1984extensions}, which states that a set of $2 N_T$ vectors in 
%a $N_s$-dimensional Euclidean space 
$\mathbb{R}^{N_s}$ can be mapped into a randomly generated subspace $\mathbb{R}^{\ell}$ of dimension $\ell \sim \mathcal{O} \left(\log (2 N_T ) \,/\epsilon^2\right)$ or greater, while preserving all pairwise $L_2$ norms to within a factor of $(1\pm \epsilon)$, where $0<\epsilon<1$, with a very high probability. 
Here, $\epsilon$ is the 
%admissible 
mismatch or distortion tolerated in the  pairwise $L_2$ norms between any two filters after projection. Thus, RP also approximately preserves any  statistic of the dataset that is characterized by such pairwise distances. 
The RP of $\bf H$ onto $\mathbb{R}^{\ell}$
%a $k$-dimensional subspace 
produces 
%${\bf \tilde{H}} \equiv \bf {H \Omega}$;
$\bf {H \Omega}$;
%, whose components are
%$ \tilde{H}_{2N_T\,k} = \sum_{N_s} \,H_{2N_T\,N_s} ~{\Omega}_{N_s\,k}$.
%
the accuracy of this data-oblivious transformation depends on the target dimensions and sampling distribution of the ${N_s\times \ell}$ projection matrix $\mathbf\Omega$. While it is enough to sample the entries independently and identically distributed from a sub-Gaussian distribution, here we choose them independently from   
a Gaussian distribution with mean zero and variance $1/{\ell}$, i.e., $\mathcal{N}(0,\, 1/{\ell})$, %{\color{red} (Traditionally, the second argument is the variance and not stdev. Check if 1/sqrt(k) is variance.)} 
thus producing a Gaussian quasi-orthonormal random
matrix~\cite{dasgupta1999elementary,dasgupta2000experiments} such that
$ \langle \mathbf\Omega  \mathbf\Omega^T \rangle = I$. Results
obtained from RP-based processing can vary depending on the actual
choice of the distribution (from which elements of $\mathbf\Omega$ are
drawn), and in a statistical sense, these results arising from
different choices of $\mathbf\Omega$ are expected to be equivalent due
to the quasi-orthonormality of the projection.
%CHECK: , as explained geometrically in Appendix~\ref{appendix:geometricExplanation}.
(See Supplemental Material for a geometric explanation.)

%--- Introduce RSVD here
%\noindent{\textit{\textbf{RP-based template matrix factorization.}}}
\section{RP-based template matrix factorization}
\label{rsvd_section}

\begin{figure}[htbp]
\centering{
\includegraphics[width=.5\textwidth]{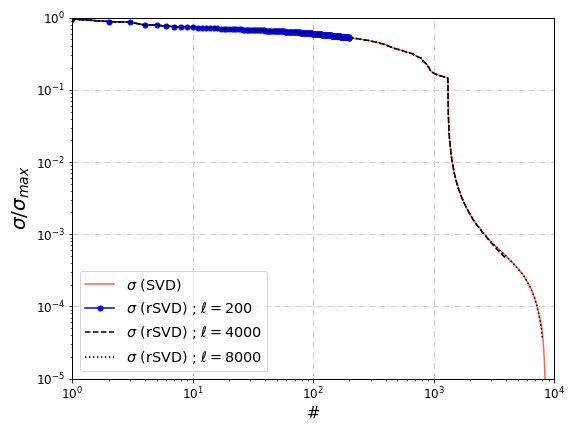}}
\caption{\label{fig:Sig_Comp} Comparison of singular values $\sigma$ for a template matrix ${\mathbf{H}}$ of size $(2 N_T \times N_s) \equiv 9130 \times 65536$, normalized by the maximum singular value $\sigma_{max}$, as obtained from SVD and RSVD factorization. RSVD is performed in target dimensions $\Rl$ where $\ell=200, 4000$ or $8000$. The spectrum of eigenvalues is seen to fall steeply. This example template bank was constructed using non-spinning signal model for component mass parameters ($m_{1,2}$) in the range $2.5 \Msun \leq m_1, m_2 \leq 17.5 \Msun$. As seen here, the top-$\ell$ eigenvalues obtained by RSVD agree very well with the spectrum obtained by traditional SVD factorization. 
}
\end{figure}

The key idea behind an $\ell$-truncated SVD approximation of ${\bf H}$ is to reconstruct the rows of the template matrix using the top-$\ell$ right-singular vectors. This approximation works well because $\mathbf{H}$ has a fast-decaying spectrum, as shown in Fig.~\ref{fig:Sig_Comp}. In making the truncation, one effectively reduces ${\bf H}$ to its $\ell$-rank approximation ${\bf H}^{(\ell)}$ ~\cite{Golub:1996:MC:248979,halko2011finding}.~\footnote{Note that ${\bf H}^{(\ell)}$ has the same dimensions as ${\bf H}$.} Further, for a bank of normalized templates, it is easy to show that the average fractional loss in SNR due to the truncation is given as $\langle\delta \rho / \rho \rangle \leq \|{\bf H}-{\bf H}^{(\ell)}\|^2_F \, / \, \|{\bf H}\|^2_F$, where $\langle \, \rangle $ denotes average over the bank of templates and $\| \cdot\|_F$ is the Frobenius norm~\cite{Golub:1996:MC:248979}. 
However, the existing SVD algorithms do not scale well with increasing dimensions and redundancies of the template database.

Randomized-SVD (RSVD)~\cite{halko2011finding} is a RP-based matrix-factorization technique to obtain an $\ell$-rank matrix factorization ${\bf H}^{(\ell)}$
%${\bf H}^{(\ell)} \approx Q_{2N_T \times \ell}\, B_{\ell\times N_s}$
such that, for some specified $\eta>0$, $\|{\bf H} - {\bf H}^{(\ell)}\|_F \le \min_{ \{ {\bf X}\,:~\rank({\bf X})\le \ell \} }\|{\bf H} - {\bf X}\|_F (1 + \eta)$ with high probability.~\footnote{Note that the values of $\eta$ and $\epsilon$ can be different.}
In one implementation, the RSVD algorithm proceeds by first projecting the individual row-vectors in the template matrix ${\bf H}$ to $\Rl$ by using ${\bf\bar \Omega}_{N_s \times \ell} \in {\mathcal{N}}(0,1/{\ell})$, thereby yielding ${\bf \bar{H}}_{2N_T \times \ell} = \mathbf{H \, \bar \Omega}$. The latter can be used to perform an SVD-like factorization directly in $\Rl$ through a series of operations like the ones described below.
In passing, we note that while $\bf \bar{H}$ is an object in a lower-dimensional Euclidean space relative to $\bf H$, it is not constituted of {\em time-decimated} templates.
%{\color{red}{$\bf \bar{H}$ is the random projection of $\bf H$ to a lower dimensional Euclidean space, therefore the rows of $\bf \bar{H}$ should not be interpreted as time-decimated template waveform.}}
 
%
% \textbf{Left vertical-axis}: Comparison of singular values $\sigma$ of the template matrix of size $9130 \times 65536$ (normalized by the maximum singular value $\sigma_{max}$) as obtained from SVD and RSVD factorization. RSVD is performed in target dimensions $\Rl$ where $\ell=4000$ or 8000. The spectrum is seen to fall steeply. This template bank was constructed using non-spinning signal model for component mass parameters ($m_{1,2}$) in the range $2.5 \Msun \leq m_1, m_2 \leq 17.5 \Msun)$. There is obviously a trade-off between the desired accuracy and the reduction in the cost and memory to carry out the factorization. As seen from the traces the agreement between the singular values of ${\bf H}^{(\ell=4000)}$ is slightly worse than those of ${\bf H}^{(\ell=8000)}$ as compared to the singular values of the full template matrix ${\bf H}$.\\
%\textbf{Right vertical axis}: Average SNR reconstruction error as
%given in Eq.~(\ref{eq:svdError}) for $\ell=8000$. As expected, the reconstruction improves with increasing number of basis vectors.

Figure~\ref{fig:Sig_Comp} compares the singular values obtained by the RP-based factorization against those from a direct SVD factorization.
As seen there,
%CHECK: repeated from above: ${\bf{H}}$ has a fast-decaying spectrum, and
it is typically sufficient to take $\ell \ll N_T$. (Since $N_T \leq N_s$,
as mentioned above, it follows that $\ell \ll N_s$ as well.) In fact, the
numerical value of $\ell$ chosen in RSVD may be smaller than the
theoretical JL bound prescribed for preserving pairwise $L_2$
distances between the $2N_T$ rows to a $\epsilon$-distortion
factor. Working with the reduced sized matrix ${\bf \bar{H}}$ leads to
significant computational savings, while producing a decomposition
that closely approximates the optimal $\ell$-rank factorization of
${\bf H}$. 
%CHECK: It is customary to oversample $\ell$ by a small factor for better numerical results. This is implicitly assumed below. 
The optimum choice of $\ell$ depends on the shape of the eigenvalue
spectrum. In the Monte-Carlo simulations presented in Supplemental Material,
%the template matrix has been compressed by a factor $\gtrsim 80$ to
%arrive at the $\ell$-rank approximation. As mentioned elsewhere, 
%For the template bank used in the Monte-Carlo simulation,
we choose $\ell =200$. The corresponding average SNR loss for a set of $500$ CBC
signals added to simulated aLIGO noise is $\langle \delta \rho /
\rho \rangle = 2 \times 10^{-4}$ in that study (see Fig.~\ref{Fig:Average_SNR_loss_rSVD} in Supplemental Material).

%===================
RSVD thus proceeds by obtaining a set of orthogonal bases for the column space of ${\bf \bar{H}}$ by using a thin-QR decomposition\cite{Golub:1996:MC:248979}: ${\bf \bar{H}} = {\bf Q}\, {\bf R}$, where ${\bf Q}$ is an orthonormal matrix with dimensions ${2N_T \times \ell}$. The approximate rank-$\ell$ decomposition is then obtained as 
${\bf {H}}^{(\ell)} = {\bf Q}\,({\bf Q}^T {\bf{H}}) = {\bf Q\,B}$, where $ {\bf B}_{\ell \times Ns} \equiv {\bf Q}^T \bf{H}$ is a  matrix that defines the orthonormal projection of the template waveforms into the compressed subspace. %{\color{red} (Check if the transposition convention here is consistent with one used in previous sections!)}
%From the factorization of ${\bf H}$ %in $\Rl$, as shown above, 
It is clear that one can use the $\ell$ rows of ${\bf B}$ as the surrogate templates, which in turn can be used to correlate against the detector data $\vec{S}$. These can be further combined with ${\bf Q}$ to reconstruct $\rho$ in $\mathbb{R}^{N_s}$. We can thus use the QB decomposition itself to improve the efficiency of both the time and frequency domain searches by constructing ${\bf H}$ appropriately, with 
templates from the corresponding domains.
 
Instead of randomly projecting the column space of $\mathbf{H}$, the method can be generalized by applying RP on both the row and column spaces~\cite{halko2011finding}. This bilateral RSVD method is particularly useful when both $N_s$ and $N_T$ are very large. 
%This bilateral RSVD approach can then be used to recover the top-$\ell$ left and right singular vectors of the template matrix in the original dimension. 

%\noindent{\textit{\textbf{Reconstruction of SNR.}}}
\section{Reconstruction of SNR}

The rank-$\ell$ matrix factorization of $\mathbf{H}$ using RSVD is given by $\mathbf{H}^{(\ell)} = {\bf QB}$. Thus, the SNR $\rho'_\alpha$, for any given $t_0$, can be reconstructed in $\RNs$ as
\begin{eqnarray}
 \rho'_{\alpha}
 &=& \left( H^{(\ell)}_{(2\alpha -1)} - i H^{(\ell)}_{(2\alpha)}\right) \vec{S}^T \nonumber \\
 &=& \sum_{\nu=1}^\ell \left ( Q_{(2\alpha -1)\nu} - i Q_{(2\alpha)\nu} \right ) \left ( B_\nu \, \vec S^T \right ).
\label{eq:snr_reconstruction}
\end{eqnarray}
Using Pythagoras theorem, and the fact that $||\mathbf{H}||_{F}^2 = 2N_T$, it is easy to show that the average fractional loss of SNR is given by 
\begin{equation}
 \left \langle \frac{\delta \rho}{\rho} \right \rangle   \leq \frac{||\mathbf{H}||_{F}^2 - ||\mathbf{H}^{(\ell)}||_{F}^2}{||\mathbf{H}||_{F}^2} = 
 %\frac{1}{2} \left ( 1 - \frac{\sum_{\mu = 1}^{\ell}{\sigma_{\mu}^2}}{2N_T} \right ).
  1 - \frac{\sum_{\mu = 1}^{\ell}{\sigma_{\mu}^2}}{2N_T} \,,
\label{eq:svdError}
\end{equation}
where 
%$\langle \, \rangle $ denotes average over the bank of templates and the 
$\sigma_\mu$ are the eigenvalues of $\mathbf{H}^{(\ell)}$.
For the example discussed in Fig.~\ref{fig:Sig_Comp}, $\sum_{\mu =
  1}^{\ell}\sigma_{\mu}^2 / (2N_T) < 1$ but approaches unity
monotonically with increasing $\ell$. The right-hand side of
Eq.~(\ref{eq:svdError}) can be calculated efficiently by evaluating the Frobenius norm of $\mathbf{B}$ directly (i.e., without
explicitly finding the eigenvalues of $\mathbf{H}^{(\ell)}$
first). Thus, the QB decomposition can indeed serve as a stand-in
replacement for the SVD factorization. (For an efficient method of 
explicitly calculating the SVD factors from the RP-based factorization
%CHECK: see Appendix~\ref{appendix:SVD_from_RP}.
see Supplemental Material.)

Ideally one would like to use $\left \langle {\delta \rho}/{\rho} \right \rangle$ as the control parameter and solve Eq.~(\ref{eq:svdError}) for the optimum value of $\ell$. However, this is a hard problem and in practice the value is set by a process of trial and error, which thankfully can be done off-line even when the computation in Eq.~(\ref{eq:snr_reconstruction}) is conducted on-line.

%As such, a 
%\noindent{\textit{\textbf{}}}
%subsection{Random projection based correlations}

A naive implementation of  matched-filter in time-domain can be very
expensive, with a complexity of $\mathcal{O}(N_s^2)$ per template for $N_s$ time-shifts. Of course, the Fast Fourier transform can reduce this to $\mathcal{O}(N_s \log N_s)$. 
It is however more efficient instead to first project the two aforementioned whitened time-series vectors in $\mathbb{R}^{N_s}$ to a random $k$-dimensional ($k\ll N_s$) subspace and then calculate
the match (using circular cross-correlations), as seen in Fig.~\ref{fig:SNR_Opt} of Supplemental Material. 
In fact, for the template part, one can directly project the rows of the ${\bf B}$ matrix
(which serve as surrogate templates) to $R^k$ ($k \leq N_s$). In this context, RP reduces the complexity of calculating the matches by a factor $N_s / k$.  
%It is however more efficient instead to first project the two aforementioned whitened time-series vectors in $\mathbb{R}^{N_s}$ to a random $k$-dimensional ($k\ll N_s$) subspace and then cross-correlate; the pairwise distance-preserving property of RP guarantees that the matched-filter output in $\mathbb{R}^k$ will be approximately equal to $\rho_\alpha$ in $\mathbb{R}^{N_s}$. 
%(see Appendix~\ref{appendix:RPCorr} on how FFT-like algorithms enable its fast computation~\cite{Golub:1996:MC:248979}). 
(See Supplemental Material for how FFT-like algorithms enable its fast computation~\cite{Golub:1996:MC:248979}.)
%On top of its usage in efficiently factorizing large template banks,
%One can also use RP to further project the rows of the ${\bf B}$ matrix (which serve as surrogate templates) and the signal to a lower dimensional Euclidean space $R^k$ ($k \leq N_s$) in order to calculate the match (circular cross-correlations), as seen in Fig. 2 of Supplemental Material. 

%\noindent{\textit{\textbf{Computational complexity analysis.}}}
\section{Computational complexity analysis}

The straightforward SVD factorization of ${\bf H}$ requires $\mathcal{O}(N_T^2\, N_s)$ floating-point operations, assuming $N_T \leq N_s$. In comparison, the cost of the %RP-based QB factorization 
RP matrix factorization is $\mathcal{O}\left ( \ell N_T N_s + (\ell^2
  N_T - \frac{2}{3}\ell^3) + \ell N_T N_s + \ell N_s \right )$. In
this last expression, we have included partial contributions from
first projecting the template matrix to $\mathbb{R}^{\ell}$, then
taking the thin-QR decomposition of $\mathbf{\bar{H}}$ using Householder's
method~\cite{numerical_recipies}, followed by the cost of constructing
$\mathbf B$ and calculating its Frobenius norm, respectively. For practical cases,
one expects $\ell \ll N_s$, due to which the cost of factorizing
$\mathbf H$ can be 
%CHECK: several 
orders of magnitude less than a full SVD
factorization. This advantage is not just realized off-line, but can
also directly impact the total on-line cost of the searches owing to a lower value of $\ell$ alone: Figure~\ref{Split_Comp} shows that for moderate sized banks one effectively ends up using $\sim 3-4$ times fewer surrogate templates in the on-line portion of the search from the new RP-based factorization. This improvement is expected to be higher for larger banks. 

%{\color{blue} Below we compare with LLOID.}
For on-line searches, the number of floating point operations per second (flops) in our method is $N_{\rm flops} = (2\ell\, k f_s + 2\ell\, N_T f_s + k\, f_s)$. The first term is the number of floating-point operations required for computing the cross-correlation between the surrogate templates (rows of ${\bf B}$) and the data vector; the second term is the cost of reconstruction of the SNR for every template; and the third term is the cost of projecting the data vector into the lower-dimensional space. In the SVD-only method, the expression for $N_{\rm flops}$ is analogous, except that instead of the last term above, it has a down-sampling cost that is similarly insignificant as our projection cost. The primary difference between the two methods is that owing to our use of RSVD and RP, $\ell$ and $k$ are less than the number of basis templates and the number of time samples of data used, respectively, in the SVD-only method. For the crucial last couple of seconds of the cross-correlation analysis for CBC signals we have evaluated that our method is an order of magnitude faster than the SVD-only method.

\section{Conclusion}

In summary, here we introduced random projection-based techniques that
hold promise for factorization of large template matrices and
cross-correlation of templates in a scalable and
computationally efficient way, which can aid more complex searches,
such as of CBCs with generic spins, and, hence, improve the chances for new discoveries.

%\noindent{\textit{\textbf{Acknowledgments.}}}
%, Various endowment funds, etc.
\begin{acknowledgments}

We would like to thank Surabhi Sachdev for carefully reading the manuscript and making useful comments. This work is supported in part by DST's SERB grants EMR/2016/007593 and  DST/ICPS/CLUSTER/Data Science/General/T-150, NSF grant PHY-1506497, and the Navajbai Ratan Tata Trust. A large set of data analysis studies were performed on the Sarathi computing cluster at IUCAA.

\end{acknowledgments}

%\clearpage

%\bibliography{randomProj_prd}
%\bibliographystyle{apsrev4-1}

%\end{document}

%%%%%%%%%%%%%%%%%%%%%%%%%%%%%%%%%%%%%%%%%%%%%%

%\clearpage

%\bibliography{randomProj_prd}
%\bibliographystyle{apsrev4-1}

%============= SUPPLEMENTAL MATERIAL ===============%

\onecolumngrid
%\hline 
\newpage

\begin{center}
  \textbf{\large 
  	Supplemental Material:\\ 
	Random projections in gravitational wave searches of compact binaries}\\[.2cm]

  	Sumeet Kulkarni,$^{1}$ Khun Sang Phukon,$^{2}$ Amit Reza,$^{3}$ Sukanta Bose,$^{4,5}$\\
	Anirban Dasgupta,$^{3}$ Dilip Krishnaswamy,$^{6}$ and Anand S. Sengupta,$^{3}$\\[.1cm]
	
  {\itshape 
  	${}^1$\IISERPUNE\\
  	${}^2$\IITK\\
  	${}^3$\IITGN\\
  	${}^4$\IUCAA\\
  	${}^5$\mbox{\WSU}
  	${}^6$\IBM\\[0.2cm]}
	%(Dated: \today)\\[0.5cm]
\end{center}

\twocolumngrid

\setcounter{section}{0}
\setcounter{equation}{0}
\setcounter{figure}{0}
\setcounter{table}{0}
\renewcommand{\theequation}{S\arabic{equation}}
\renewcommand{\thefigure}{S\arabic{figure}}
\renewcommand{\bibnumfmt}[1]{[#1]}
\renewcommand{\citenumfont}[1]{#1}

The purpose of this document is three-fold. First, we provide a
geometric explanation for why RSVD preserves the top singular
subspaces. Second, we present an efficient method for explicitly 
calculating the SVD factors from the RP-based factorization.
Third, we demonstrate how random projection can be used to
compute the match between two normalised templates directly in the target space.

%The purpose of this Appendix is three-fold. First, we provide a
%geometric explanation for why RSVD preserves the top singular
%subspaces. Second, we present an efficient method for explicitly 
%calculating the SVD factors from the RP-based factorization.
%Third, we demonstrate how random projection can be used to
%compute the SNR time-series directly in the target space.

%\noindent{\textit{\textbf{}}}
\section{Geometric explanation for top singular subspace preservation under RSVD.}
\label{appendix:geometricExplanation}

\begin{figure*}[!htbp]
\centering{
\includegraphics[width = 0.75\textwidth]{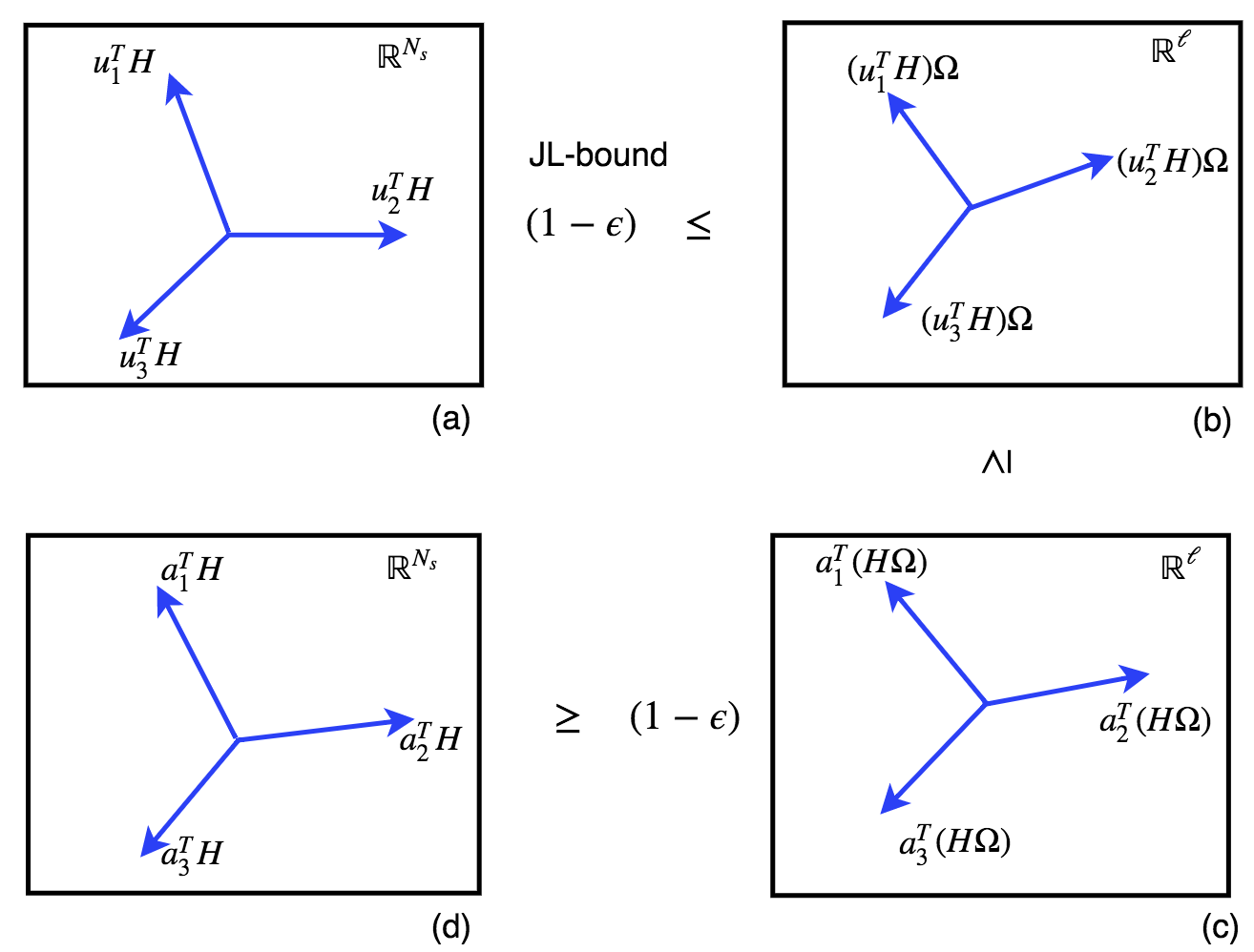}}
\caption{\label{fig:Geo_Int} Depiction of the geometrical intuition behind the preservation of the top singular subspaces in RSVD. 
Finding the top-$k$ singular vectors for a given ${\bf H}$ is akin to the finding the orthogonal directions $u_1, \ldots u_k$
such that the sum of lengths of the vectors $u_i^T {\bf H}$ is maximized. 
Subfigure~{\textbf{(a)}} shows these vectors in the original feature space. 
Subfigure~{\textbf{(b)}} depicts the action of random projection of these vectors to a lower-dimensional space using the projection matrix $\bf \Omega$.
The length-preserving properties of random projections guarantee that the distortion in their lengths lies within a factor of $1\pm \epsilon$.
Subfigure~{\textbf{(c)}} depicts the action of $\bf{H \Omega}$ along its top-$k$ left singular directions $\{a_1, \ldots, a_k\}$, such that  $\sum_{i\leq k} \|a_i^T ({\bf H}{\bf \Omega})\|^2$ is maximum. Note that these vectors $a_i^T ({\bf {H \Omega}})$, can also be geometrically interpreted as the the random projection of the vectors $a_i^T {\bf H}$ (as shown in subfigure~{\textbf{(d)}} using $\bf \Omega$. 
Through this chain of subfigures, it follows that the low rank approximation $\bf A A^T H$ found by RSVD captures most of the energy in the optimal rank-$k$ approximation $\bf U U^T H$.
} %end_caption
\end{figure*}

Here we provide a geometric explanation behind the statement that RSVD preserves the top singular subspaces. Figure~\ref{fig:Geo_Int} depicts this intuition.

%for how RSVD works. Figure \ref{fig:Geo_Int} depicts the geometrical intuition behind the claim that RSVD preserves the top singular subspaces. 
We denote the $L_2$ norm of a vector as $\|\cdot\|$. 
Recall that given ${\bf H}$, finding out the top-$k$ singular vectors is akin to the question of finding out orthogonal directions $u_1, \ldots u_k$
such that if ${\bf U} = [u_1, \ldots, u_k]$,  then
\begin{equation}
\label{Eq:Frob_rel1}
\|{\bf UU}^T {\bf H} \|_F^2 = \|{\bf U}^T {\bf H}\|_F^2 = \sum_{i=1}^k \|u_i^T {\bf H}\|^2
\end{equation}
is maximized (squared lengths of vectors in subfigure (a)). 

Given the length-preserving properties of random projection,
the sum of the squared lengths of $u_i^T {\bf H}$ is equivalent, up to an approximation factor of $1\pm \epsilon$, to the sum of squared lengths of $(u_i^T {\bf H}){\bf \Omega}$ ((subfigure (b)), i.e., 
$$
(1 - \epsilon)\sum_{i\le k} \|u_i^T {\bf H}\|^2  \le \sum_{i\le k}  \|(u_i^T {\bf H}){\bf \Omega}\|^2.
$$ 
Hence, we instead find orthonormal vectors $\{a_1, \ldots, a_k\}$ that
are the top-$k$ left singular directions of ${\bf {H\Omega}}$, such that $\sum_{i\leq k} \|a_i^T ({\bf H}{\bf \Omega})\|^2$ is maximum (subfigure (c)), thus achieving $\sum_{i\leq k} \|a_i^T ({\bf H\Omega})\|^2 \ge \sum_{i\leq k} \|u_i^T ({\bf H\Omega})\|^2$.

Again by using the length preservation of random projection, we have that 
$\|a_i^T {\bf H}\|^2 \ge  (1 - \epsilon)\|(a_i^T {\bf H})\bf \Omega\|^2 $ (subfigure (d)). 
Putting the above steps together, we find
\begin{eqnarray}
\label{Eq:Ineq1}
\sum_{i=1}^{k} \|a_i^T {\bf H}\|^2 &\ge& (1- \epsilon)\sum_{i=1}^k
\|(a_i^T {\bf H}){\bf \Omega}\|^2 \nonumber \\
&\ge&  (1- \epsilon) \sum_{i=1}^k \|(u_i^T {\bf H}) {\bf \Omega}\|^2  \nonumber \\
&\ge&  (1- \epsilon)^2\sum_{i=1}^k \| (u_i^T {\bf H})\|^2\,.
\end{eqnarray}
Recall that by definition of singular vectors, we already have 
\begin{equation}
\label{Eq:Ineq2}
\sum_{i=1}^{k} \|a_i^T {\bf H}\|^2 \le  \sum_{i=1}^{k} \|u_i^T {\bf H}\|^2.
\end{equation}
Combining Eqs.~(\ref{Eq:Ineq1}) and (\ref{Eq:Ineq2}), one can obtain the following inequality: 
\begin{equation}
\label{Eq:Ineq3}
\big(1-\epsilon^2 \big) \leq \frac{\sum_{i=1}^{k} \|a_i^T {\bf H}\|^2}{\sum_{i=1}^{k} \|u_i^T {\bf H}\|^2} \leq 1\,.
\end{equation}
Let $\bf{A}$ $=[a_1,\ldots,a_k]$ and recall that $\bf{U}$ $=[u_1,\ldots,u_k]$. Given that $\bf{U}$ is an orthonormal matrix, the matrix  $\bf{UU}^T \bf{H}$ represents the projection of columns of $\bf{H}$ onto the subspace spanned by the columns of $\bf{U}$, and hence is a rank-$k$ approximation of $\bf{H}$.
Using an argument similar to that used in Eq.~(\ref{Eq:Frob_rel1}) it is possible to obtain
\begin{equation}
\label{Eq:Frob_rel2}
\|\bf{AA}^T \bf{H}\|_F^2 = \sum_{i = 1}^{k}{\|a_i^T \bf{H}\|^2}\,.
\end{equation}
Therefore, using Eqs.~(\ref{Eq:Frob_rel1}), (\ref{Eq:Ineq3}) and (\ref{Eq:Frob_rel2}), it is clear that the low rank approximation $\bf{AA}^T \bf{H}$ found by RSVD captures most of the energy in the optimal rank-$k$ approximation $\bf{UU}^T \bf{H}$.

%Therefore, it follows from the above argument,  that the low rank approximation $\bf{AA}^T \bf{H}$ found by RSVD captures most of the energy in the optimal rank-$k$ approximation $\bf{UU}^T \bf{H}$. 

%\noindent{\textit{\textbf{}}}
\section{Calculating the SVD factors from the RP-based factorization
efficiently}
\label{appendix:SVD_from_RP}

%The current GW pipeline uses the singular values of $H$ to estimate the fractional loss  of SNR due to the rank-$\ell$ approximation. 
An efficient method of explicitly calculating the SVD factors from the RP-based factorization is now presented. This is intended as a bridge between the two methods. Singular values can be obtained by performing an SVD on ${\bf B}$, or by first calculating ${\bf T}_B = {\bf B}{\bf B}^T$, of size $\ell\times \ell$. The eigenvectors ${\bf U}_{T_B}$ of ${\bf T}_B$ are identical to the left-singular vectors of ${\bf B}$, and the eigenvalues ${\bf \Sigma}_{T_B}$ are equal to the squares of the singular values of ${\bf B}$. As ${\bf T}_B$ is a much more compressed matrix compared to ${\bf B}$, it is far more efficient to store it in memory and factorize it thereby revealing the singular values and left-singular vectors.  These in turn can be further used to calculate the singular values of ${\bf H}^{(\ell)}$.
%From the $QB$ decomposition of $H$, it is also possible to recover 
The top-$\ell$ right-singular vectors of ${\bf H}$ in %
%the original feature space (i.e. in 
$\mathbb{R}^{N_s}$ can be obtained using the left-singular vectors of ${\bf T}_B$:  ${\bf U}_{H}^{(\ell)} \approx {\bf Q}\, {\bf U}_{T_B}$. Similarly, it can be trivially checked that ${\bf  \Sigma}_H {\bf  V}_H^T = {\bf  U}_{T_B}^T\, {\bf  B}$. Thus, all the pieces of the SVD factorization of ${\bf H}$ can be recovered from RSVD factors, but at a small fraction of the computational cost of the former. In doing so, the advantages of RP-based factorization can be directly transferred to the current SVD-based data analysis pipelines.

\section{Random projection based correlations}
\label{appendix:RPCorr}

 A naive implementation of  matched-filter in time-domain can be very expensive, with a complexity of $\mathcal{O}(N_s^2)$ per template. Here we show 
% repeated from before: where $N_s$ is the number of time-points or dimension of a template.
how random projections can be applied to reduce this cost considerably: 
the whitened time-series vectors, in the form of the template and the data, in $\mathbb{R}^{N_s}$ can be first projected to a random $k$-dimensional ($k\ll N_s$) subspace and then cross-correlated; the pairwise distance-preserving property of RP guarantees that the matched-filter output $\rho^\prime_\alpha$ in $\mathbb{R}^k$  will be approximately equal to $\rho_\alpha$ in $\mathbb{R}^{N_s}$, i.e.,
%\begin{eqnarray}
%\langle \rho^\prime_\alpha \rangle &=& 
%\langle \left ( H_{(2 \alpha -1)} \, \mathbf{\Omega}  - \iu H_{(2\alpha)} \, \mathbf{\Omega} \ri%ght ) \left ( \vec{S} \, \mathbf{\Omega} \right ) ^T \rangle \\
%&=& \langle \left ( H_{(2 \alpha -1)} \, \mathbf{\Omega} \mathbf{\Omega}^T  - \iu H_{(2\alpha)} \, \mathbf{\Omega}\mathbf{\Omega}^T \right ) \vec{S}^T \rangle \nonumber \\
%&=& \rho_\alpha \nonumber \, ,
%\end{eqnarray}
\begin{equation}
\langle \rho^\prime_\alpha \rangle =
\left\langle \left ( H_{(2 \alpha -1)} \, \mathbf{\Omega}  - \iu H_{(2\alpha)} \, \mathbf{\Omega} \right ) \left ( \vec{S} \, \mathbf{\Omega} \right ) ^T \right\rangle = \rho_\alpha \nonumber \, ,
\end{equation}
%where we have used $ \langle \mathbf{\Omega} \mathbf{\Omega}^T \rangle = {I}$ (by construction). 
because $ \langle \mathbf{\Omega} \mathbf{\Omega}^T \rangle = {I}$.

\begin{figure}[!htbp]
\centering{
\includegraphics[width = .5\textwidth]{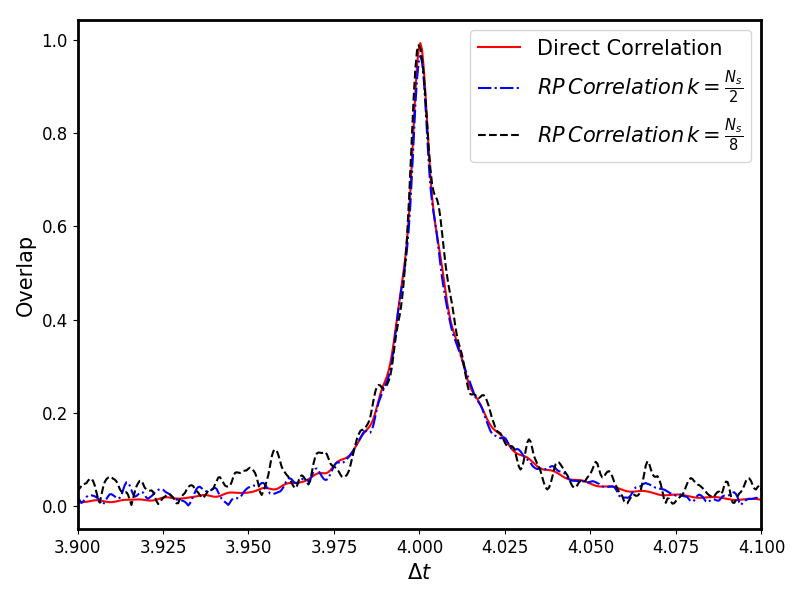}}
%%%%%  Old figurefr_rp_rsvd_compare.eps}
\caption{\label{fig:SNR_Opt} The phase-maximized overlap time series
  of a normalized template (corresponding to equal component masses
  $m_{1,2} = 6~\Msun$)  correlated against a copy of itself time
  shifted by $4$ seconds. The template is taken to be $8$ seconds long
  and sampled at $2048$ Hz. The circular correlations are calculated
  both \emph{directly} in a high dimensional space $\mathbb{R}^{N_s}$
  and using RP in a lower dimensional Euclidean space
  $\mathbb{R}^{k}$,
as discussed in Sec.~\ref{appendix:RPCorr}; 
here $N_s= 16384$ and $k=N_s/2, N_s/8$. The agreement between the two traces shows that RP can be used to efficiently calculate overlaps in a lower dimensional space.}
%The SNR time series obtained by calculating the correlation of $\vec S$ against a template in $\RNs$ (blue solid line) using PyCBC, where $N_s = 65\, 536$ is compared to the reconstructed time series using the RP based methods presented here.  In the latter cases, the template matrix of $2N_T = 23\,040$ rows is at first RSVD factorized %in $\mathbb{R}^\ell$ for $\ell = 570$.  The red colored dashed line depicts the  SNR time series, reconstructed from the cross-correlations  between $\vec S$ and the resulting basis vectors in $\RNs$.   Furthermore, correlations with $\vec S$ are calculated in the target space $\mathbb{R}^k$ for $k=4\,000$ and the reconstructed time series is represented by the green line. The  reconstruction error in SNR in the red-dashed and green lines are $6\times 10^{-6}$ and $8\times 10^{-3}$ respectively.   The component mass parameters of the injected signal ($m_1 = 1.78 M_\odot, m_2 = 1.7\,M_\odot$) are chosen to be the same as one of the templates in the bank.  The $\vec S$ contains the last 10 seconds of waveform for the mass parameters, buried in 16 seconds of gaussian noise. The distortion factor on SNR from the JL-lemma is $\epsilon=0.1$. }
\end{figure}

It can also help searches to use the RP-based correlation in conjunction with the RP-based QB factorization of the template matrix described above.
The key to this fusion between the two RP-based methods lies in the
fact that instead of matched-filtering the data $\vec S$ against every
template in the bank, one can use the reduced set of $\ell \leq 2N_T$
row vectors of $\mathbf B$ as surrogate templates for this
purpose. These correlations can be calculated in $\mathbb{R}^{k}$ by
projecting $\vec S$ and each row vector $B_\nu\,\in\,\mathbb{R}^{N_s}$
to the target $k$-dimensional subspace. Such projections preserve the
inner products between the data and $B_\nu$ at every relative
time-shift within the $\epsilon$ bound, as guaranteed by the JL
lemma. The complex SNR for each template can be reconstructed using
the coefficient matrix $\bm{\mathcal{C}}_{N_T\times\ell}$, whose
elements are $\mathcal{C}_\alpha \equiv \left ( Q_{(2\alpha-1)\nu} -
  \iu Q_{(2\alpha)\nu} \right )$ as shown in the main text.
%Eq.~(\ref{eq:snr_reconstruction}). 
The phase-maximized SNRs of the templates are given by the  modulus of resulting complex SNRs. 

The construction proceeds as follows: Suppose 
data are sampled at a rate $f_s$ and that the duration in which one decides to search for the signal's time of arrival is $\tau$, which is taken to be longer than the longest template in the bank. Then the number of points over which one is discretely searching for $t_0$ is $f_s\tau$. 
%In order to search for CBC signals over the time of arrival parameter, we 
We next construct a partial circulant matrix ${\bf K}(B_\nu)$ for every row of ${\bf  B}$, such that its dimensions are $(f_s\tau) \times (f_s\tau + N_s)$.
%Its every row has $N_s$ elements, with the
Its $n^{\rm th}$ row $K_n(B_\nu)$ 
%has is a zero-padded, being a 
is a copy of $B_\nu$ that is time-shifted by an amount $\Delta t_n = \, n/f_{s}$,
%, where $f_{s}$ is the sampling frequency, 
where $n\in [0, (f_s\tau) ]$.
%over the admissible range of time-shifts. 
The remaining $f_s\tau$ elements in each row are set to zero.
The data vector $\vec S$, with $(f_s\tau+N_s)$ time-points, and the
circulant matrices can both be randomly projected to the subspace
$\mathbb{R}^k$ using $\mathbf{\Omega}$. Their subsequent multiplication 
%of the random-projected circulant matrix with the projected data vector 
is used to construct
%a matrix of output vectors of 
the cross-correlation:
\be
\rho^\prime_{\alpha} (\Delta t_n) = \sum_{\nu}\mathcal{C}_{\alpha\,\nu}\, \left ( K_n (B_\nu) {\bf\ \Omega} \right ) \, (\vec{S} {\bf \Omega})^T  \,,
\label{eq:rpcorrelation}
\ee
where $\alpha$ is the index over the templates in the bank %(i.e. a row of $\mathbf{H}$), 
$\nu = 1, \ldots, \ell$ is the index on the rows of ${\bf  B}$, and ${\bf \Omega}$ has dimensions of $(f_s\tau+N_s) \times \ell$.
%$S^{(k)},\mathcal{T}^{(k)}\,\in\,\mathbb{R}^k$ represent the projected data vector and projected partial circulant matrices, respectively. 
A circulant matrix can be diagonalized using FFT-like algorithms to enable efficient processing of matrix-vector products involving such matrices~\cite{Golub:1996:MC:248979}. Figure~\ref{fig:SNR_Opt} compares the phase-maximized overlap computed using this method for two choices of $k$ with that computed directly, i.e., without employing random projections.

\begin{figure}[!htbp]
\centering{
    \includegraphics[width=.5\textwidth]{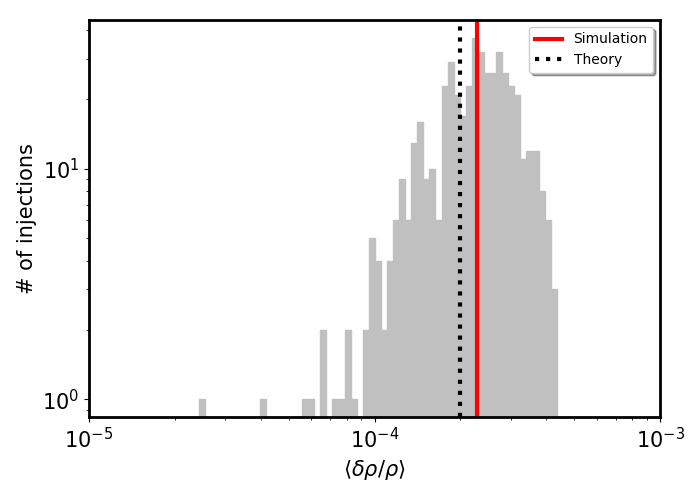}
    }
\caption{\label{Fig:Average_SNR_loss_rSVD}   
The distribution of averaged SNR loss summarizing the results of the Monte-Carlo injection study presented in Sec.~\ref{appendix:MonteCarlo}. The mean of the distribution (solid red vertical line) is close to the theoretically expected mean loss (dotted black vertical line). The similarity of the results of this study with a similar study presented in Ref. \cite{Cannon:singular} establishes the validity of efficient RP-based QB factorization of the template banks presented in this work.}
%This figure shows how SNR is lost in the stage \RNum{1} of our RP-based search method. In the left panel of figure, a scatter plot which compares the reconstructed SNRs  in the $\RNs$-space with the SNRs computed using PyCBC is shown. The color bar represents fractional loss in SNRs ($\delta\rho/\rho$), relative to SNRs computed with PyCBC. In middle panel, the histogram of $\delta\rho/\rho$ for the exact-match injections is plotted. In the right panel, the histogram of fractional loss of SNRs for the non-exact-match injections  are shown.}
\end{figure}
%

%========== Appendix D =========%
\section{SNR reconstruction using QB decomposition of the template bank}
\label{appendix:MonteCarlo}

We have shown above that very large template banks can be efficiently QB decomposed using the RSVD algorithm thereby representing the template matrix by its rank-$\ell$ approximation. One can reconstruct the SNR time series for each template in this bank to a high degree of accuracy by projecting the data on the top $\ell$ basis vectors, akin to the truncated SVD paradigm. The average SNR loss can be estimated from the singular values corresponding to the discarded basis vectors. 

We now present results from a Monte-Carlo study to explicitly
demonstrate that the fractional SNR loss (averaged over the bank) due
to the rank-$\ell$ approximation of the template matrix closely
follows the theoretically estimated value, as evaluated using the
expression for $\delta \rho / \rho$ in Sec.~IV of main text, thereby 
validating the accuracy of the RSVD factorization.

We consider a template bank $\mathbf{H}$ containing $N_T = 581$ templates covering the component mass space: $5 \leq m_{1,2}/\Msun \leq 15$ using non-spinning TaylorT4 waveforms. 
For this study, each waveform was taken to be $8$ seconds long, sampled at $2048$ Hz, thereby setting $N_s = 16384$. 
%We use the canonical advanced LIGO aLIGOZeroDetHighPower design power spectral density. 
We use the aLIGO Zero Detuned High Power (ZDHP) noise power spectral density~\cite{aLIGO_ZDHP}.
Signals were simulated for $500$ CBC sources, with component masses randomly chosen from the aforementioned mass range. These were separately added to colored Gaussian noise with aLIGO ZDHP power spectral density. The amplitudes of the injected signals were adjusted for a target SNR of 8.
The mass parameters of most of these signals were different from those
of the templates in $\mathbf{H}$.

%Simulated advanced LIGO data was generated using signal from $500$ injections with random component masses over this range to which coloured Gaussian noise was added. The amplitude of the injected signal was adjusted for a target SNR of 8. 

The template matrix was first QB decomposed to a rank $\ell=200$ approximation using the RSVD algorithm that corresponded to an averaged SNR loss $\langle \delta \rho / \rho \rangle = \frac{\sum_{\ell + 1}^{2 N_T}  \sigma_i^2}{\sum_{1}^{2 N_T} \sigma_i^2} = 2 \times 10^{-4}$. This
threshold was decided based on the spectrum of the singular values of
$\mathbf{H}$, which was observed to fall sharply -- similar to the
examples shown in Fig. 2. of the main text.

For each simulated signal injection, the SNR for every template was reconstructed using the $200$ basis vectors and compared with the SNR calculated from a direct circular correlation of these templates against the noisy data containing the injection. Thereafter the averaged SNR loss was evaluated. The distribution of this quantity over the set of all injections is shown in Fig.~\ref{Fig:Average_SNR_loss_rSVD}. As shown there, the mean of the distribution agrees well with the target set at $2 \times 10^{-4}$. Note that the correlations were computed in $\mathbb{R}^{N_s}$. 

A similar study was presented by Cannon et al.~\cite{Cannon:singular} using truncated SVD factorization of the template matrix. The similarity of the results establishes the validity of efficient QB decomposition of large template banks after random projection.

\bibliography{randomProj_prd}
\bibliographystyle{apsrev4-1}

\end{document}